\documentclass[sigconf]{acmart}

\usepackage{booktabs} 
\usepackage[linesnumbered, ruled,vlined]{algorithm2e}
\usepackage{amsmath}

\begin{document}
\title{Open Domain Question Answering Using Web Tables}

\author{Kaushik Chakrabarti}
\affiliation{Microsoft Research}
\email{kaushik@mirosoft.com}

\author{Zhimin Chen}
\affiliation{Microsoft Research}
\email{zmchen@microsoft.com}

\author{Siamak Shakeri}
\affiliation{Microsoft (Bing)}
\email{siamaks@microsoft.com}

\author{Guihong Cao}
\affiliation{Microsoft (Bing)}
\email{gucao@microsoft.com}

\begin{abstract}
Tables extracted from web documents can be used to directly answer many web search queries.
Previous works on question answering (QA) using web tables have focused on factoid queries, i.e.,
those answerable with a short string like person name or a number.
However, many queries
answerable using tables are non-factoid in nature. In this paper, we develop
an open-domain QA approach using web tables that works for both factoid and non-factoid queries.
Our key insight is to combine deep neural network-based semantic similarity between the query
and the table with features that quantify the dominance of the table in the document as well as the quality of the information in the table.
Our experiments on real-life web search queries show that our approach
significantly outperforms state-of-the-art baseline approaches.
Our solution is used in production in a major commercial web search engine
and serves direct answers for tens of millions of real user queries per month.
\end{abstract}


\maketitle

\section{Introduction}\label{sec:intro}

The Web contains a vast corpus of HTML tables. In this paper, we focus on one class
of HTML tables called ``relational tables'' \cite{cafarella:vldb08,cafarella:vldb09,venetis:vldb11}.
Such a table contains several entities and their values
on various
attributes, each row corresponding to an entity and each column
corresponding to an attribute.
We henceforth refer to such tables as web tables.

\begin{figure}[t]
\vspace{-0.4cm}
\begin{center}
\includegraphics[width=3.4in]{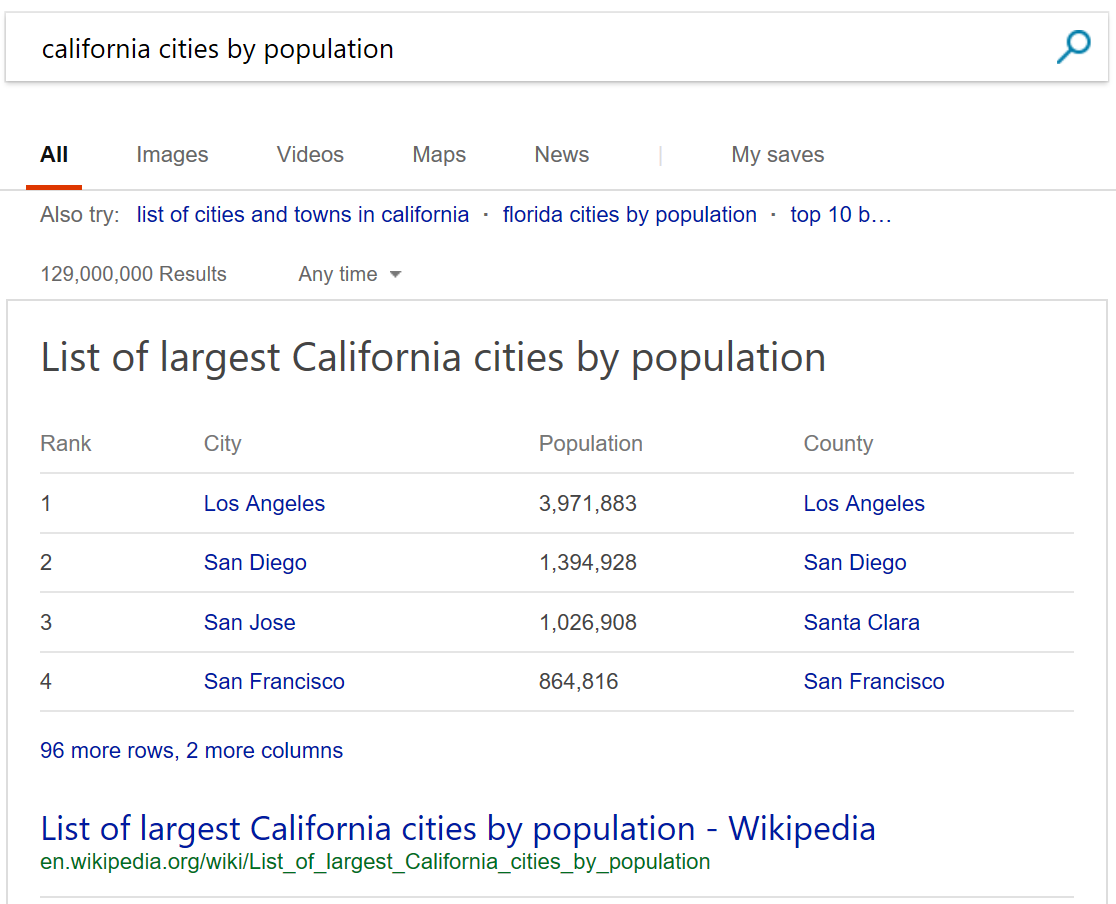}
\vspace{-0.15in}
\caption{\small \label{fig:taexample} Example of table answer in a major commercial search engine}
\end{center}
\vspace{-2em}
\end{figure}

A web table is the best way to directly answer several important classes of web queries: (1) 
those seeking the set of
entities that satisfy certain criteria (e.g., `california cities by population') (2)
those looking for the top entity or entities based on certain ranking criteria (e.g., `richest actor in the world') (3)
those seeking the value of an entity on a particular attribute (e.g., `galaxy s7 display size')
or (4) those looking for information that is naturally tabular
like concert schedule of an artist (e.g., `dale ann bradley schedule')
or roster of a team (e.g., `maryland terrapins roster') or specs of a device (e.g., `galaxy s8 specs') or sports statistics of a player/team (e.g., `babe ruth stats').
A web table returned as an answer to such a query is henceforth referred to as a \emph{``table answer''}.
Figure \ref{fig:taexample} shows a table answer on a major commercial search engine.
Identifying queries that can be answered using a table and finding the table answer for them is an important challenge.

Previous works on question answering (QA) using web tables
focus on factoid questions, i.e., ones answerable with a short string like a person name or a number.
They focus on returning the correct cell of the table that answers the question \cite{pasupat:acl2015,sun:www2016}.
However, many queries that are answerable using tables
are non-factoid in nature. For example, all queries listed above except `galaxy s7 display size' are non-factoid queries.
Non-factoid queries cannot be answered using a single table cell;
the answer is either the entire table or a subset of rows and columns in the table.
For example, in Figure \ref{fig:taexample}, the answer is the entire table.
We develop an approach that works for \emph{both factoid and non-factoid queries}.

We adopt the architecture used in information retrieval (IR)-based passage answering systems 
\cite{pascabook:2003,jurafskybook:qachapter}
and adapt it for table answers. We first use a web search engine to identify a pool of candidate tables
and then identify the best table answer, if one exists, from among them.
We refer to the latter step as
the \emph{table answer selection}.
Performing table answer selection with high precision and reasonably good coverage is hard.
This is main technical challenge we tackle in this paper.

\noindent \textbf{Baseline approaches:} We consider two baselines, IR-based approach adapted from 
passage answering systems
and dominating table approach. \\
\noindent $\bullet$ \emph{IR-based approach}: The main idea is to
compute the IR-style ``match'' between
the query and each candidate table and select the table answer based on it.
They also extract
the desired answer type from the question and the main entity type(s) present in each candidate table
and compute the degree of match between them.
A high degree of IR match together with a strong type match indicates a high likelihood of being a good answer.
While this approach works well for passage answers, it performs poorly for table answers.
Consider the query `graph database'. The top result returned by the search engine is the Wikipedia article on graph database
(\url{en.wikipedia.org/wiki/Graph_database}). It contains a table containing the list of all
well-known graph database software.
The above baseline will identify it as the table answer
as it occurs in the top document and has perfect IR match with the query.
However, it is not a good answer as the user is most likely interested in learning what a
graph database is, i.e., she is interested in the non-table content (text paragraphs in this case) of that document.
Since our candidate space does not include non-table content, it is \emph{difficult to ensure
that the chosen table answer is better than the non-table content in top documents}.
\\
\noindent $\bullet$ \emph{Dominating table approach}: 
We consider an alternate approach, referred to as the dominating table approach,
to avert the above problem. We observe that there
are many documents where a table (or multiple tables) ``dominates'' the document, i.e.,
it occupies a significant fraction (say, more than half) of the overall document content.
Suppose such a document is returned as a top result by the search engine.
Assuming that the top-ranked document(s) perfectly answers the query,
the dominating table should be a good answer to the query.
For example, for the query `california cities by population',
the table in Figure \ref{fig:taexample} dominates the top document and is a good answer.
This approach also has its pitfalls.
The top document
may not always perfectly answer the query (the perfect document may be ranked lower or simply may not exist).
If it happens to contain a dominating table, it may be erroneously returned as the answer.
Consider the query `upcoming races in saratoga ny'.
Both major US commercial search engines
returns the document \url{www.runningintheusa.com/race/List.aspx?State=NY} on top
which contains a dominating table listing all the upcoming races in the entire NY state
(which may or may not contain races in Saratoga). The perfect document (the one that contains a list of races in Saratoga)
\url{localraces.com/saratoga-springs-ny}
is ranked lower. The above approach will incorrectly return the NY state table as the answer.
Another common situation is that the top document does contain the ideal table answer but it is not formatted as a HTML table.
If there is a dominating table in a lower ranked, non-perfect document, it may be erroneously returned as the answer.
Assume that the search engine returns the Saratoga document on top
and the other document in second position.
Since the list in the Saratoga document is not formatted as a HTML table, it will not be extracted.
So, this approach will still return the NY state table as the answer.



\noindent \textbf{Key insights:} Our main insight is to use \emph{both IR match between the query and the table
and the degree of dominance of the table relative to the document to select the table answer}.
We convert the task to a supervised classification problem and incorporate both IR-match
and table dominance as features.
Using both of them avoids the pitfalls of using IR match alone (first baseline)
or table dominance alone (second baseline).
The graph database table, which erroneously qualified as a table answer in the first baseline,
will no longer qualify because although it has a strong IR match, the degree of dominance is low. 
The NY state table, which erroneously qualified as a table answer in the second baseline, 
will also not qualify because although the table dominates the document,
the IR match is weak (no match for 'Saratoga').
Furthermore, concepts are often expressed using different vocabularies and language styles in tables and queries,
hence word-level matching models like TF-IDF and BM25
can be inadequate for computing query-table match \cite{huang:cikm13,shen:cikm14,tan:arxiv15}. 
To address this issue, we leverage deep neural network-based latent semantic models previously proposed for query-document matching
to compute query-table match \cite{huang:cikm13,shen:cikm14}.



\noindent \textbf{Contributions:} Our research contributions can be summarized as follows:
\\
\noindent $\bullet$ We study open domain\footnote{\small{Since the queries can be from any domain}} QA
using web tables that works for factoid as well as non-factoid queries.
To the best of our knowledge, this is the first paper to study table answers for both types of queries.
\\
\noindent $\bullet$ We introduce the table answer selection problem. We propose a novel approach that combines deep neural network-based semantic similarity between the query and table
with features that quantify the dominance of the table relative to the document as well as quality of information in the table.
\\
\noindent $\bullet$ We perform extensive experiments on real-life search queries on a major commercial web search engine.
Our approach significantly outperforms the state-of-the-art baselines approaches.
Our solution is used in production in a major web commercial search engine and 
serves direct answers for tens of millions of real user queries per month.

We present the system architecture and define the table answer selection
and table answer classification problems in Section 2.
We describe the features used for table answer selection in Sections 3 and 4. 
Section 5 describes the snippet generation algorithm. Section 6 presents the experimental evaluation.
Section 7 discusses related work and Section 8 concludes the paper.

\section{System Architecture and Problem Statement} \label{sec:archi}

We first describe the system architecture.
We then formally define the technical problems we study in this paper.
We conclude by outlining the key steps of our solution.

\begin{figure}[t]
\vspace{-0.4cm}
\begin{center}
\includegraphics[width=3.5in]{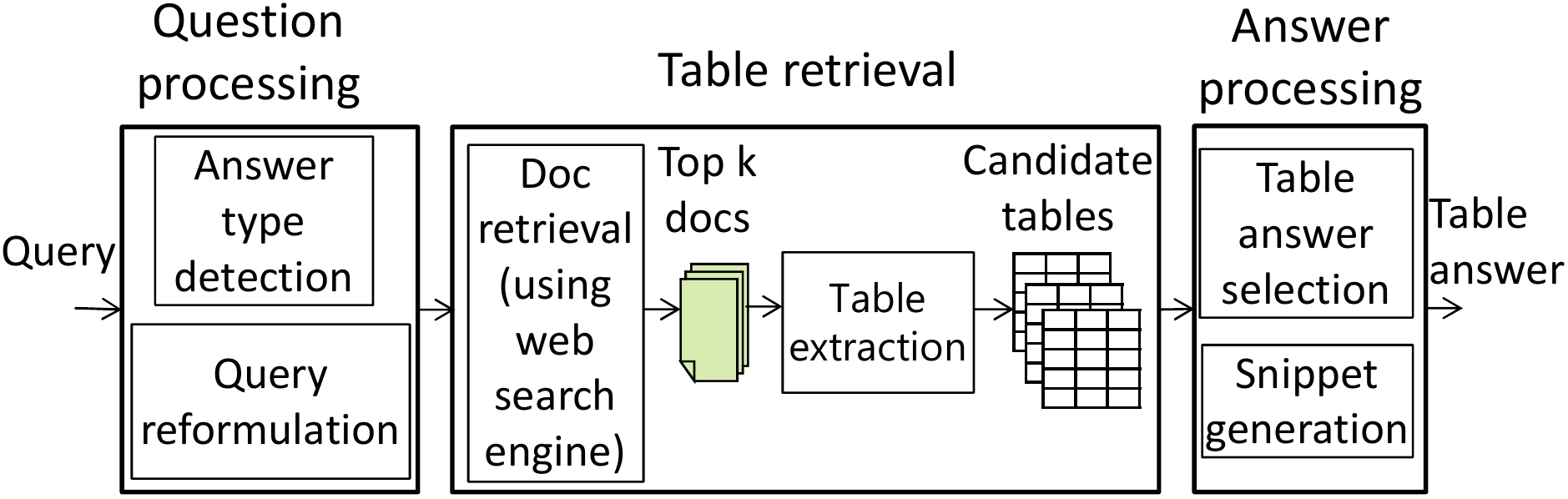}
\caption{\small \label{fig:archi} System architecture}
\end{center}
\vspace{-0.6cm}
\end{figure}

\subsection{System Architecture}
We adopt the architecture used in IR-based passage answering systems \cite{pascabook:2003,jurafskybook:qachapter}
and adapt it for table answers. Figure \ref{fig:archi} shows the system architecture.
It consists of 3 components:
\\
\noindent \textbf{Question Processing}: This component extracts useful information from the query
like the desired answer type. 
It also determines whether to reformulate the query before sending it to a web search engine.
For this component, we adopt the techniques developed in the context of passage answering \cite{pascabook:2003,jurafskybook:qachapter}.
\\
\noindent \textbf{Table retrieval}: This component sends the query (either original or reformulated) to the web search engine,
retrieves the top $k$ ($k \approx 5-10$ ) documents and extracts the tables occurring in them. We refer to them as \emph{candidate tables}. It is important to consider documents beyond the first position
as the best answer is sometimes present in a lower ranked document. At the same time, we do not consider beyond $k=10$ documents
as it leads to too many candidate tables and increases the cost of table answer selection.
We adopt the techniques previously proposed to extract and classify relational tables \cite{cafarella:relWT,wang:www02}.
For each table, we extract the following information: (i) url, title and h1 heading of the document
(ii) heading of the section/subsection the table belongs to (typically, h2/h3/h4 headings)
(iii) text immediately preceding the table (iv) caption (content of <caption> tag) and header/footer rows if they exist
(v) column names if they exist and (vi) all the cell values.
We refer to (i)-(v) as \emph{metadata} of the table
and (vi) as the \emph{cell content}.
Furthermore, each row in a
relational table corresponds to an entity and there is
typically has one column that contains the names of those entities \cite{venetis:vldb11}.
For example, each row in the table shown in Figure \ref{fig:taexample} corresponds to a city
and the second column from the left contain that names of those cities.
This column is referred to as the subject column of the table.
We identify the subject column of the table using techniques similar to \cite{venetis:vldb11}.
\\
\noindent \textbf{Answer processing}: This component has two subcomponents.
The first one performs table answer selection, i.e., determines whether any of the candidate tables answers the query and, if yes, identifies that table. The second one 
computes a ``snippet'' of the table to be displayed on the search engine result page (Figure \ref{fig:taexample} shows an example).
The former is the technical focus of the paper, the latter is described briefly in 
in Section \ref{sec:snippet}.

\subsection{Problem Statement}
We formally define the technical problem we address in this paper.
\begin{definition}[Table Answer Selection Problem]
Given a query $Q$ and a pool $\mathcal T$ of candidate tables,
determine whether there exists any candidate table that answers the query.
If yes, return the candidate table $T_{best} \in \mathcal T$ that best answers the query, 
else return $\{\}$.
\end{definition}

Our approach is to first solve the table answer classification problem which is defined as follows:
\begin{definition}[Table Answer Classification Problem]
Given a query $Q$ and a table $T$, return the score $F(Q,T)$
which represents the degree to which $T$ is an answer for $Q$.
\end{definition}

\begin{table}
\begin{center}
\vspace{-0.1in}
\begin{tabular}{|p{0.05in}p{3.0in}|}
\hline
1 & For each candidate table $T \in \mathcal T$, compute $F(Q,T)$ by invoking the table answer classifier. \\
2 & Pick the candidate table $T_{best} = argmax_{T \in \mathcal T} F(Q,T)$ with the highest classifier score \\
3 & If $F(Q,T_{best}) > \theta$, return $T_{best}$, else return $\{\}$. \\
\hline
\end{tabular}
\caption{Selecting table answer based on table classification results}
\label{tab:tas}
\end{center}
\vspace{-0.2in}
\end{table}

Subsequently, we perform the table answer selection using the algorithm shown in Table \ref{tab:tas}.
It invokes the table answer classifier for each candidate table; this is feasible
as we typically have a small set of candidate tables. It then
picks the one $T_{best}$ with the highest classifier score.
If $T_{best}$'s score exceeds a threshold $\theta$, we return it as the answer, otherwise we return no answer.
The choice of $\theta$ allows the system to obtain the desired precision and recall trade-off.

\subsection{Key Steps of Our Solution}

The main technical challenge is to build a table answer classifier
that accurately predicts the score $F(Q,T)$.
This is the
focus of the rest of the paper.
Like building any machine learning classifier, it consists of 3 steps: \\
\noindent (i) \textbf{Feature creation}: We create features that helps to discriminate between
good and bad table answers. The main novelty of our solution lies in this step. 
As discussed in Section \ref{sec:intro}, we create two groups of
features: the ones that compute the match between the query and table
and ones that compute the dominance of the table relative to the document
as well as quality of information of the table. 
We refer to the two groups as \emph{query-table similarity features}
and \emph{table dominance and quality features} respectively.
We describe them in Sections \ref{sec:qtsim} and \ref{sec:tableimp} respectively.
\\
\noindent (ii) \textbf{Learning algorithm}: We use an off-the-shelf learning algorithm (LogitBoost boosted tree learner) for this purpose.
\\
\noindent (iii) \textbf{Training process}: We obtain training data and train and test the model.
We discuss this in detail in Section \ref{sec:expts}.

\section{Query-Table Similarity Features}\label{sec:qtsim}

We describe the first group of features, namely query-table (Q-T) similarity features.
Instead of inventing new models for computing Q-T similarity,
we map the table into one or more ``documents''
and leverage previously proposed query-document similarity models 
to compute Q-T similarity.
We first present the techniques to map a table to documents
and then describe the query-document similarity models we leverage.

\subsection{Mapping Table to Documents}

\begin{figure}[t]
\vspace{-0.4cm}
\begin{center}
\includegraphics[width=3.5in]{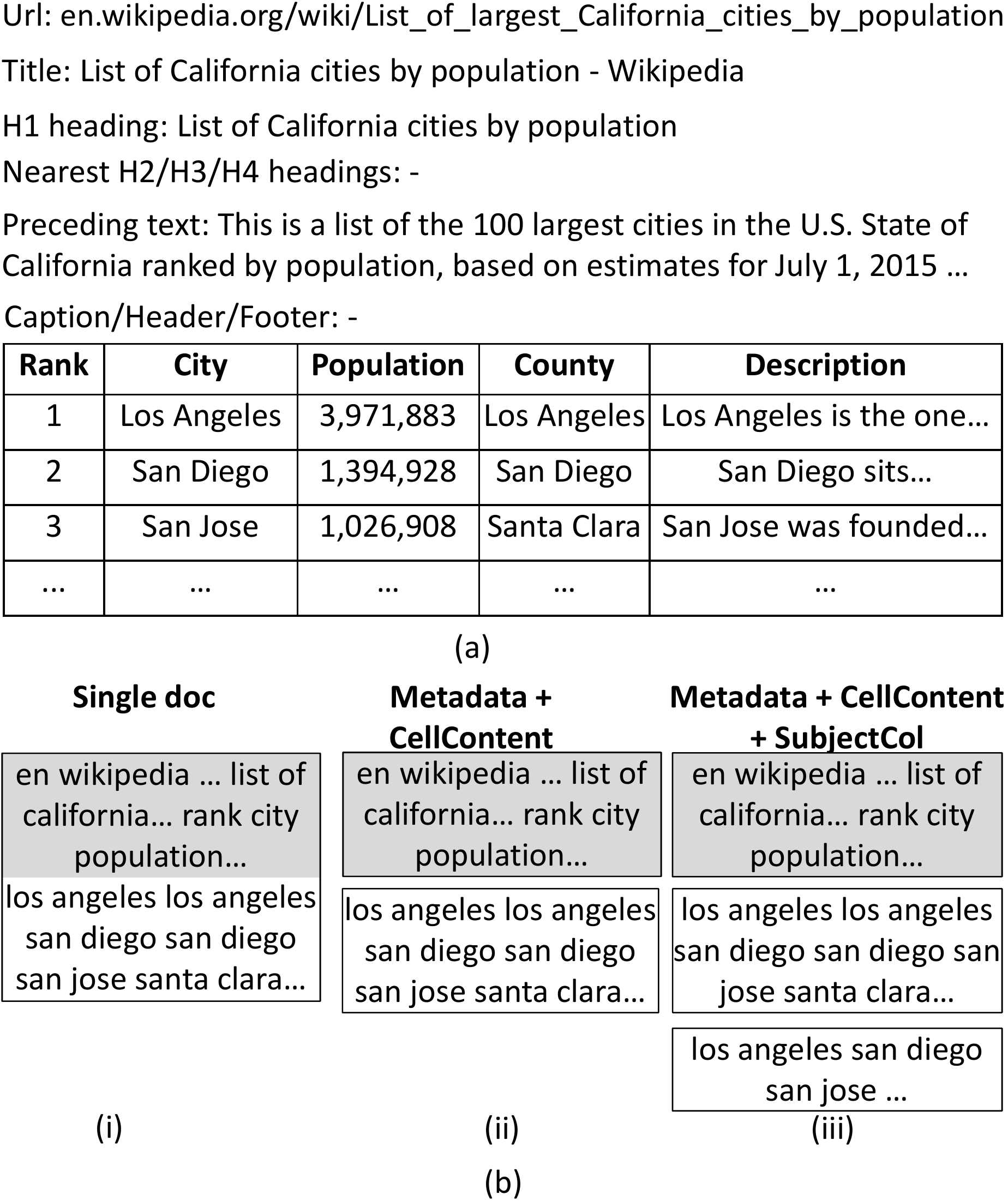}
\caption{\small \label{fig:mapping} Different techniques to map table to documents}
\end{center}
\vspace{-0.6cm}
\end{figure}

We need to construct document(s) from the table such that existing query-document similarity models
can capture the desired match
between the query and the table contents. For some Q-T pairs, we need to match with only the metadata of
the table (i.e., pagetitle, pageheading, column names, etc.).
Figure \ref{fig:mapping}(a) shows the metadata and cell contents
of the table shown in Figure \ref{fig:taexample}. For the query `california cities by population',
we need to match with only the metadata. Now consider the queries `santa clara cities by population'
or 'san jose population'. The same table is a good table answer for such queries as well.
However, matching with metadata alone will not help selecting this table
as there is no match of the keywords `santa clara' and `san jose'
respectively in the metadata.
In such cases, we need to match with the cell contents as well.
We study several different ways to construct the documents with the above requirements in mind;
they are illustrated in Figure \ref{fig:mapping}(b).
\\
\noindent $\bullet$ \textbf{Single document}: Since match with both metadata and cell contents is important,
we construct a document (denoted by $Doc(T)$) by concatenating the contents of all the metadata fields as well as the cell values.
Figure \ref{fig:mapping}(b)(i) shows the constructed document for the example table in Figure \ref{fig:mapping}(a).
Specifically, we concatenate the tokens in the url, page title, h1 heading of the page, the headings of the section/subsection
the table belongs to, content of the <caption> tag if present, header/footer rows if present
and names of the columns if present. For the cell content, we concatenate the cell values row by row starting from the top row;
the cells inside each row are concatenated from left to right. We skip columns with numeric values since search queries typically do
not contain numbers.
For example, in Figure \ref{fig:mapping}(b)(i), we skip the `Rank' and `Population' columns.
\\
\noindent $\bullet$ \textbf{Metadata + CellContent}:
The main limitation of the single document approach is that
it cannot distinguish between query keyword matches in the metadata fields
vs those in cell contents. 
This can lead to erroneous similarity scores.
Consider the query `us state capitals'
and the two candidate tables in \url{en.wikipedia.org/wiki/List_of_capitals_in_the_United_States}:
one containing the current capitals (perfect answer) and the other containing the historical
capitals (bad answer). The first table has better match with metadata
while the second one has better match with cell contents (as the keyword 'capital' has many matches in the cell contents).
The single document approach 
ends up choosing the latter (as it contains more matches in total) due to its inability to make the distinction.
We propose to construct two documents for any table $T$: a metadata document, denoted by $MDoc(T)$, by concatenating the contents of all the metadata fields and a cell document, denoted by $CDoc(T)$ by concatenating the cell values as before.
Figure \ref{fig:mapping}(b)(ii) shows the two documents for the example table in Figure \ref{fig:mapping}(a).
Subsequently, we compute separate match features for the two documents
and can make the distinction discussed above.
\\
\noindent $\bullet$ \textbf{Metadata + CellContent + SubjectCol}:
The match between the desired answer type in the query
and the main entity type in the candidate table is an important feature
in QA systems. Recall from Section \ref{sec:archi}
that there is a column, referred to as the subject column,
that contains the names of main entities in the table ((e.g., second column from the left in the table in Figure \ref{fig:mapping}).
To further distinguish matches with these entities from entities in other columns, 
we propose a construct a third document (denoted by $SDoc(T)$) containing
the column name and entities in the subject column of the document.
Figure \ref{fig:mapping}(b)(iii) shows the three documents for the example table in Figure \ref{fig:mapping}(a).

\subsection{Query-Document Semantic Similarity}

We now discuss the query-document similarity models we leverage to compute Q-T similarity.
The obvious models to leverage are the 
word-level matching models like TF-IDF and BM25.
However, due to the vocabulary discrepancies between queries and tables, those models
are inadequate for capturing the semantic similarity between a query and a table \cite{huang:cikm13,shen:cikm14,tan:arxiv15}.
We leverage two previously proposed models to compute semantic similarity between the query and the document(s) constructed from the table:
the convolutional deep semantic similarity model \cite{shen:cikm14}
and translation models\cite{Gao10,Gao:sigir11}. 

\begin{figure}[t]
\vspace{-0.4cm}
\begin{center}
\includegraphics[width=3.2in]{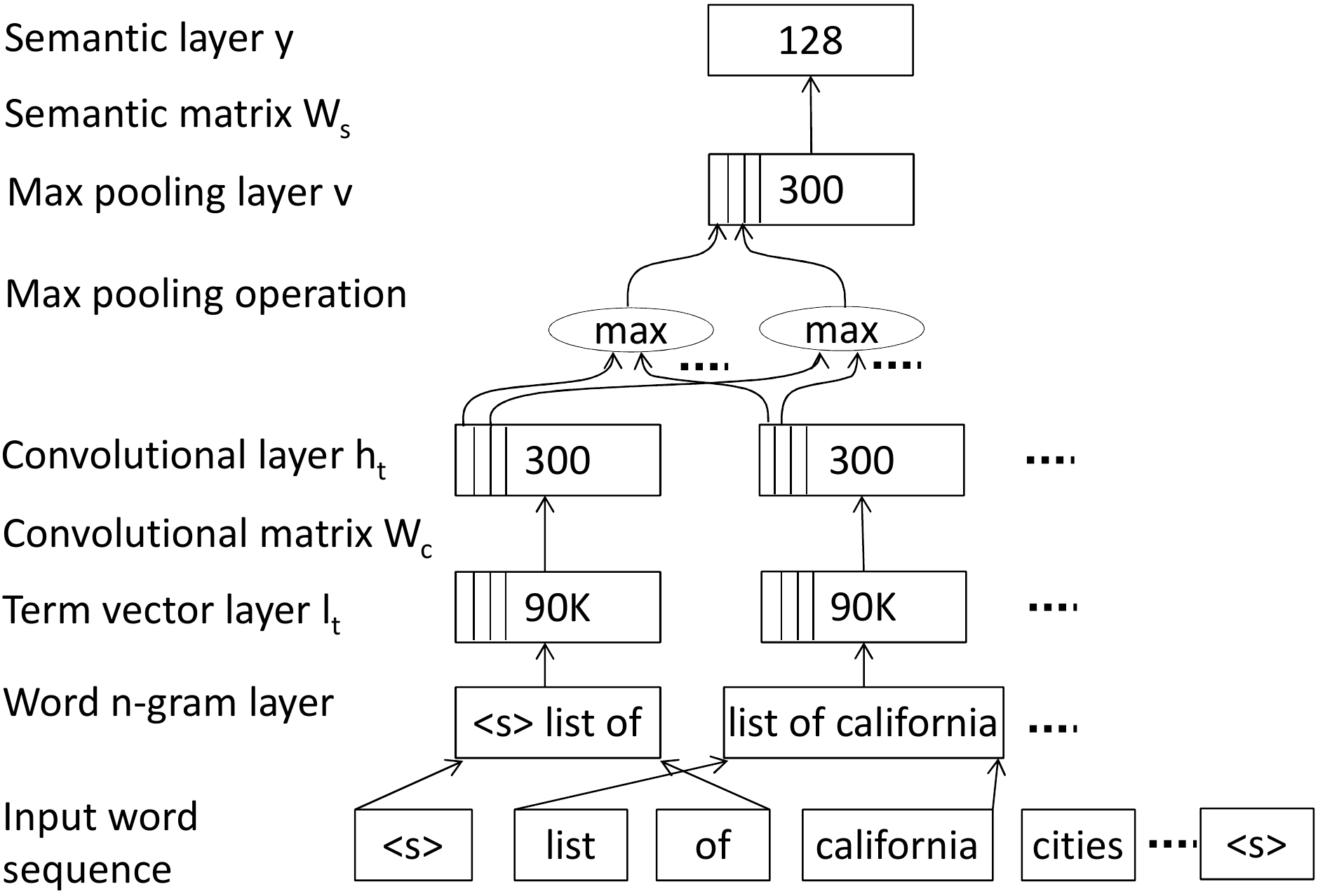}
\caption{\small \label{fig:cdssm} C-DSSM model architecture}
\end{center}
\vspace{-0.6cm}
\end{figure}

\subsubsection{Convolutional Deep Semantic Similarity Model (C-DSSM)}
This model, proposed in \cite{shen:cikm14}, uses a deep neural network to project the query and the documents to a common semantic space.
The semantic similarity is subsequently calculated as the cosine similarity between them in that space.
We choose to leverage this model as it has been shown to capture important contextual information
in the projected semantic representation.
We briefly describe the model for the sake of completeness and then the features we compute based on them.

The architecture of the model is shown in Figure \ref{fig:cdssm}.
It consists of (i) a word n-gram layer obtained by running a contextual sliding window
over the input word sequence (query or document) 
(ii) a term vector layer that transforms each word n-gram $t$ into a letter n-gram frequency vector $l_t$ 
(iii) a convolutional layer that projects the term vector $l_t$ to a local contextual feature vector 
$h_t =  tanh(W_c \cdot l_t)$ (iv) a max pooling layer 
that extracts the most salient local features to form a fixed-length sentence-level
feature vector and (v) a non-linear feed-forward neural network layer
which outputs the final semantic vector $y=tanh(W_s \cdot v)$.
The model parameters are learnt using training. Further details can be found in \cite{shen:cikm14}.
In this work, we use pretrained models that have been used for passage answering in a commercial search engine.

For any document-to-table mapping approach, we compute the CDSSM-based semantic similarity between the query and each of the constructed documents. For example, for the Metadata + CellContent + SubjectCol approach,
there are 3 constructed documents, so we compute the 3 similarities:
\begin{align}
\begin{split}
    sim1_{CDSSM}(Q,T) &= cosine(y_Q,y_{MDoc(T)}) \\
    sim2_{CDSSM}(Q,T) &= cosine(y_Q,y_{SDoc(T)}) \\
    sim3_{CDSSM}(Q,T) &= cosine(y_Q,y_{CDoc(T)})
\end{split}
\end{align}
where $y_Q$ denote the semantic vector of the query and $y_{MDoc(T)}$, $y_{SDoc(T)}$ and $y_{CDoc(T)}$
denote the semantic vectors of the metadata, subject column and cell documents for the table.
All these similarities are included as features for the classifier.

\subsubsection{Translation Models}
An alternate approach to compute semantic similarity is to leverage
translation models proposed in \cite{Gao10,Gao:sigir11}.
The model learns word translation probabilities $P(q|d)$ based on clicked
query-document pairs.
Subsequently, the semantic similarity $sim_{WTM}(Q,D)$ between a query and a document
is computed as the probability $P(Q|D)$ of $D$ being translated into $Q$.
Further details can be found in \cite{Gao10,Gao:sigir11}.

As with CDSSM, we compute the translation model-based similarity between the query and each of the constructed documents. For example, for the Metadata + CellContent + SubjectCol approach, we compute the 3 similarities:
\begin{align}
\begin{split}
sim1_{WTM}(Q,T) &= P(Q|MDoc(T)) \\
sim2_{WTM}(Q,T) &= P(Q|SDoc(T)) \\
sim3_{WTM}(Q,T) &= P(Q|CDoc(T))
\end{split}
\end{align}
All of them are included as features for the classifier.



\section{Table Dominance and Quality Features}\label{sec:tableimp}


We now describe the second group of features, namely table dominance and quality features
(\emph{table features} in short).
We further divide them into three subgroups which we describe in the three subsections.

\subsection{Fraction of Document Occupied by Table}
The fraction of the document content occupied by the table is a strong indicator of the degree of dominance
of the table relative to the document.
Larger the fraction, higher the degree of dominance.
A simple way to compute the fraction is based on the html source; take the ratio of the number of characters in
the html source inside the <table> </table> tags to that in the html source of the entire document.
While the above fraction is a useful indicator, it often does not capture the fraction
visually occupied by the table when rendered on the browser. 
For example, consider
this document about Daytona
500 race results
(\url{www.nascar.com/en_us/monster-energy-nascar-cup-series/standings/results/2015/daytona-500.raceResults.qualifying.html}).
Although the
main table on the document visually dominates the document, the source-based fraction is only 9\%.

A common reason is that the html source of the document includes
script, css styles and comments. Such invisible elements contribute to a large fraction of html source
and cause the source-based fraction to be much lower than the visual fraction.
We address this by removing script, style and comment tags before computing the size. 
With this improvement, 
the fraction of the above table increases to 32\% which is a more accurate estimate of the visual fraction. However, it still underestimates the visual fraction.

We observe that most web documents have 
a container containing the main content of the document
and several sections (header, footer, sidebar) for various advertising or navigational
purposes. The fraction of content under the <table> node versus
the root node (usually a <div> node) of the main content container is a more reliable
estimate of the visual fraction. 
However, it is often difficult to identify the main content container
as it is usually specified by a site specific style class. 
We address this challenge as follows.
Often, the main content section of a document starts with a <h1> heading. 
We use the least common ancestor <div> node of the <h1>
and <table> nodes as the root node of the main container
and compute the fraction based on it.
With this improvement, the fraction increases to 67\% 
which is more accurate estimate of the visual fraction.
We use all the above fractions as features in the classifier.

\subsection{Position of Table In Document}
The relative position of the table in the document is also an indicator of its relative importance within the document.
The author of the page tends to put important content near the top of the document, so a table 
located near the top may be the most important content
of a document even if the fraction occupied is not very high.
Consider again the Wikipedia page containing the list of capitals
in the United States (\url{en.wikipedia.org/wiki/List_of_capitals_in_the_United_States}).
The table (the one containing the current capitals) located near the top occupies only 23\% of the page
but is the most important content of the document.
As before, a simple way to compute the relative position is based on the html source.
It is the ratio of the number of characters in the html source
till the <table> tag at the start of the table
to that in the html source in the entire document. 
However, this ratio has the same limitations
as the source-based fraction: it does not always capture 
the relative position of the table as seen visually on the rendered page.
We compute new relative positions based on the improvements discussed
in the context of fraction computation. 
Furthermore, in order to differentiate between the tables based on position when there are multiple tables in a page,
we use the index of the table in the page as a feature.

\subsection{Quality of Information In Table}
Like query-independent scores computed for web documents (e.g., PageRank),
we compute the ``goodness'' of a table as a table answer \emph{irrespective of the query}.
A good table should provide correct information about entities on important attributes.
We capture that goodness via 
a set of syntactic properties of the table.
For example, a table with no or very few empty cells is better than
a table with many empty cells. Or a table with column names
is better than one without column names. Besides the above two,
we also compute number of rows and columns, presence of numeric columns, 
percentage of distinct values
on the subject column, 
and several properties capturing consistency of types of value (e.g., pure string vs pure numbers vs string containing digits) 
in columns.
We use all of the above properties as features in the classifier.

\section{Snippet Generation}\label{sec:snippet}

Suppose the table answer selector has chosen a table as the answer to the query. 
We cannot display the entire table on the search result page as it would take too much screen real estate.
Instead, we display a $m \times n$ snippet of the table on the search result page.

\begin{definition}[Table Snippet]
Consider a table $T$. Let $\mathcal R_T$ and $\mathcal C_T$ denote the set of data rows and columns of $T$.
A $m \times n$ snippet of $T$ is a table consisting of a subset $SR \subseteq \mathcal R_T$
of $m$ rows and a subset $SC \subseteq \mathcal C_T$
of $n$ columns.
\end{definition}

We display the snippet along with the names of the chosen columns as well the title/h1 heading
and the url of the document. For example, Figure \ref{fig:taexample} shows a $4 \times 4$ snippet. 
The values for $m$ and $n$ are pre-fixed based on
the device screen size (e.g., desktop vs tablet vs phone) and are typically either 3 or 4.

A good snippet should contain the answer the user is looking for so that she does not need to click
and to go to source document.
This may not be possible for many non-factoid queries. For example, for the query in Figure \ref{fig:taexample},
the snippet cannot show all of California's cities. In such cases, the snippet should show the  
entities and attributes most relevant to the user.
The goal of the snippet generation component in Figure \ref{fig:archi} is to choose the ``best'' 
$m \times n$ snippet based on the above criteria.

Since web tables are constructed for human consumption, the rows and columns in a web table are already
ordered; they are ordered based on a criteria  
that the author of the table deems important. For instance, the rows in the table in Figure \ref{fig:taexample} are ordered
in descending order of the city's population. 
Hence, unless there is evidence that some rows or columns are specifically relevant
to the query, 
we generate the snippet consisting of (i) the top $m$ rows and (ii) the leftmost $n$ columns
while making sure to include the subject column and skipping columns with 
too many empty cells/repeated values. 

\begin{algorithm}[t]
\caption{Snippet generation algorithm}
\label{alg:snipgen}
\small
Initialize selected rows $srows \gets \emptyset$\ and selected columns $scols \gets \emptyset$;

$EC \gets $ cells in entity column with keyword match ordered by desirability;

$AC \gets $ cells in other columns with keyword match ordered by desirability;

$CN \gets $ column names with keyword match ordered by desirability;

\Repeat{$srows$ and $scols$ both full or $EC$, $AC$, $CN$ are all empty}
{	
    $a \gets PopHead(EC)$;

	$SR \gets Union_m(SR, RowIndex(a))$;

	$SC \gets Union_n(SC, ColIndex(a))$;

	$b \gets PopHead(AC)$;

	$SR \gets Union_m(SR, RowIndex(b))$;

	$SC \gets Union_n(SC, ColIndex(b))$;

	$c \gets PopHead(CN)$;

	$SC \gets Union_n(SC, ColIndex(c))$;
}
\normalsize
\end{algorithm}

The most common scenarios for choosing specific rows or columns are: (i) query is about a specific
attribute (or attributes) of a specific entity in the table (e.g., query `san jose population' in Figure \ref{fig:taexample}) 
(ii) query is a filter on the table
(e.g., query `cities by population in santa clara county') (iii) query asks for specific attributes (e.g., 
query `cities by population'). 
We identify the above scenarios based on keyword matches.
If a keyword match occurs within entity column, it is
scenario (i). If a keyword match occurs within other columns, it is scenario (ii).
Finally, if a keyword match occurs within column names, it is scenario (iii). In such cases, 
those rows and columns are promoted into snippet. 

The above heuristics have some caveats.
For instance, for the query `cities by population', the row ``Salt Lake City'' should not
be promoted into snippet just because there is a keyword match on `city'. We safeguard 
against such bad cases by enforcing additional conditions on what constitutes a keyword match in the 
table data cells. First, the matched keyword needs to be \emph{``exclusive''}, i.e.,
it does not occur in metadata fields. 
If a metadata field
has the keyword `city' in it, we ignore the matches for `city' in the table data cells. 
Second, the matched keywords need to cover a significant part of the data cell content, otherwise we ignore those matches. 
This avoids the spurious keyword matches in
cells with long text.
For each data cell, we count keyword matches that satisfy both conditions
and compute a ``desirability'' score to be included
in the snippet; the desirability score is simply the ratio of the keyword match count versus number of tokens in the cell.
The snippet generation algorithm 
picks cells
in descending order of their desirability score until we have fully populated the snippet.
In computing the keyword matches, we include synonyms 
but synonym matches are discounted against exact matches
while computing the desirability scores.
Furthermore, we round robin between scenarios (i), (ii) and (iii)
to hedge the risk of misunderstanding query intent.

Algorithm \ref{alg:snipgen} outlines the snippet generation algorithm. 
Given a set $X$ and an element $y$, $Union_m(X, {y})$ is a small variant of the set union operation and is defined as follows:
\begin{align*}
Union_m(X, {y}) &= X \cup y \mbox{ if } |X| < m \\
                &= X \mbox{ otherwise }
\end{align*}
$RowIndex(c)$ and $ColIndex(c)$ of a cell $c$ denotes the row index and column index respectively.
$EC$, $AC$ and $CN$ denotes the matches based on scenarios (i), (ii) and (iii) respectively
and the algorithm picks among them in a round robin fashion.

\section{Experimental Evaluation}\label{sec:expts}

We present an experimental study of the techniques proposed in the paper. The goals
of the study are:
\\
\noindent $\bullet$ To evaluate the quality of proposed approaches
for table answer classification and table answer selection
\\
\noindent $\bullet$ To compare with the baseline approaches, i.e.,
ones using Q-T similarity features only and table features only
\\
\noindent $\bullet$ To evaluate the impact of using semantic similarity over
word-level matching
\\
\noindent $\bullet$ To evaluate the impact of various table to document mapping techniques
\\
\noindent $\bullet$ To evaluate the impact of various groups of table features
\\
\begin{table}
\vspace{-0.1in}
\begin{center}
\begin{tabular}{|l|c|c|c|}
\hline
& \textbf{Train}  & \textbf{Dev} & \textbf{Test} \\ \hline
\# Queries  & 25393 & 6987 & 4528 \\ \hline
\# Queries with at least 1 +ve Q-T pair & 15671 & 4234 & 1715 \\ \hline
\# Q-T pairs  & 43423 & 12473 & 7168 \\ \hline
\# +ve Q-T pairs & 17944 & 4887 & 3118 \\ \hline
\end{tabular}
\end{center}
\caption{Statistics of training and evaluation sets}
\label{tab:datastats}
\vspace{-0.2in}
\end{table}

\subsection{Experimental Setup}
\noindent \textbf{Datasets for Training and Evaluation:}
Previous works on QA using web tables, specifically \cite{pasupat:acl2015} and \cite{sun:www2016}, use the \textsc{WikiTableQuestions} (developed in \cite{pasupat:acl2015}) and \textsc{WebQuestions} (developed in \cite{berant:emnlp2013})
respectively. However, those datasets focus on factoid queries.
Since our system is for both factoid and non-factoid queries,
those datasets are not suitable for training and evaluation of our techniques.

We develop a new dataset using the query logs of a major commercial search engine.
We aggregated most recent 2 years of query logs and retained the queries with at
least 50 impressions.
We first identify queries that are answerable using tables.
For each query, we take the top 3 documents returned by the search engine
and check whether any of them is dominated by one or more tables, i.e., the table(s)
occupies more than 40\% of the document.
We found that about 2\% of queries satisfy this criteria.
We take a random sample of 10000 queries
from the above set to create our training and evaluation queryset.
For each query in the queryset, we obtain the candidate tables by extracting
tables from the top $5$ documents returned by the search engine.
This produces 47063 query-table pairs.
Note that we sample at the query level (not query-table pair level)
since we evaluate both the table answer classifier (which is evaluated at the Q-T pair granularity)
and the table answer selector (which is evaluated at the query granularity).

We send each pair to at least 3 experienced human judges and ask them to label each query-table pair
as either 'Good'(1) or 'Bad'(0).
If the table answers the query, is well-formed and comes from a trustworthy source,
the label should be 'Good'.
Otherwise, it should be 'Bad'.
Once the judges reach an agreement for a pair,
we save the pair and that label (referred to as the true label).

\begin{table}
\begin{center}
\vspace{-0.1in}
\begin{tabular}{|p{2cm}|p{6cm}|}
\hline
Dimension  & Techniques \\ \hline
Approach  & Q-T Sim only, Table features only, \textbf{Q-T Sim+Table} \\ \hline
Query-Doc Sim Measure & BM25, BM25+TM, \textbf{BM25+TM+CDSSM}  \\ \hline
Content used & Metadata only, CellContent only, \textbf{Metadata+CellContent} \\ \hline
Table-to-Doc Mapping & Single doc, \textbf{MDoc+CDoc}, MDoc+SDoc, MDoc+CDoc+SDoc \\ \hline
Table Features  & Frac only, Position only, Quality only, \textbf{All} \\ \hline
\end{tabular}
\end{center}
\caption{Techniques we compare in our experiments along different dimensions}
\label{tab:variants}
\vspace{-0.2in}
\end{table}

We found an imbalance in the positive and negative samples in the above data:
only 25\% of the samples are positive.
Imbalanced training data hurts the performance of the boosted tree learner
we use. To mitigate the problem, we identify queries
that are more likely to be table-answerable queries.
As discussed in Section \ref{sec:intro}, queries looking for a list
of entities (e.g., `california cities by population') or top ranked entities
(e.g., `richest actor in the world') are typically answerable by tables.
In the former case, the query contains the desired entity type in plural (e.g., cities)
while in the latter case, it contains a superlative adjective (e.g., `richest') and the desired entity type in singular (e.g., actor).
We identify such queries using a taxonomy of entity types and a list
of superlative adjectives.
We sample 5000 queries from this set, generate the query-table pairs (16001 pairs)
and collect judgments as before.
We add them to the previous set and obtain
a total of 63064 query-table pairs.
This improved the positive-negative balance to 41\% positive samples.
We randomly select 70\% of the queries 
for training, 20\% for development and 10\% for test.
We show the statistics in Table \ref{tab:datastats}.

\begin{figure*}
\vspace{-0.06in}
\begin{minipage}{0.47\linewidth}
\centering
 \includegraphics[width = 3.2in,clip]{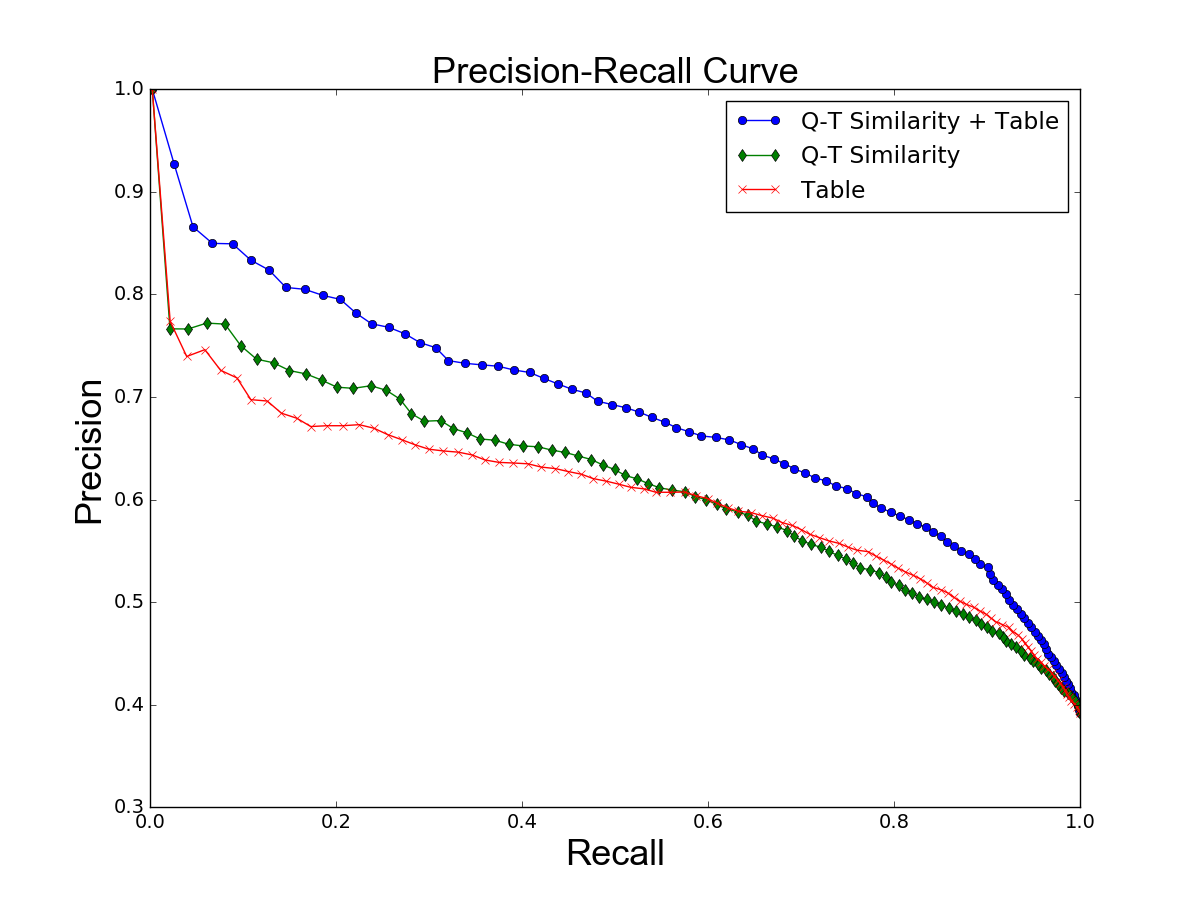}
 \vspace{-0.15in}
 \caption{\small Comparison of different approaches wrt classification performance} \label{fig:CompareApproaches1}
\end{minipage}
\hspace{0.05\linewidth}
\begin{minipage}{0.47\linewidth}
\centering
 \includegraphics[width = 3.2in,clip]{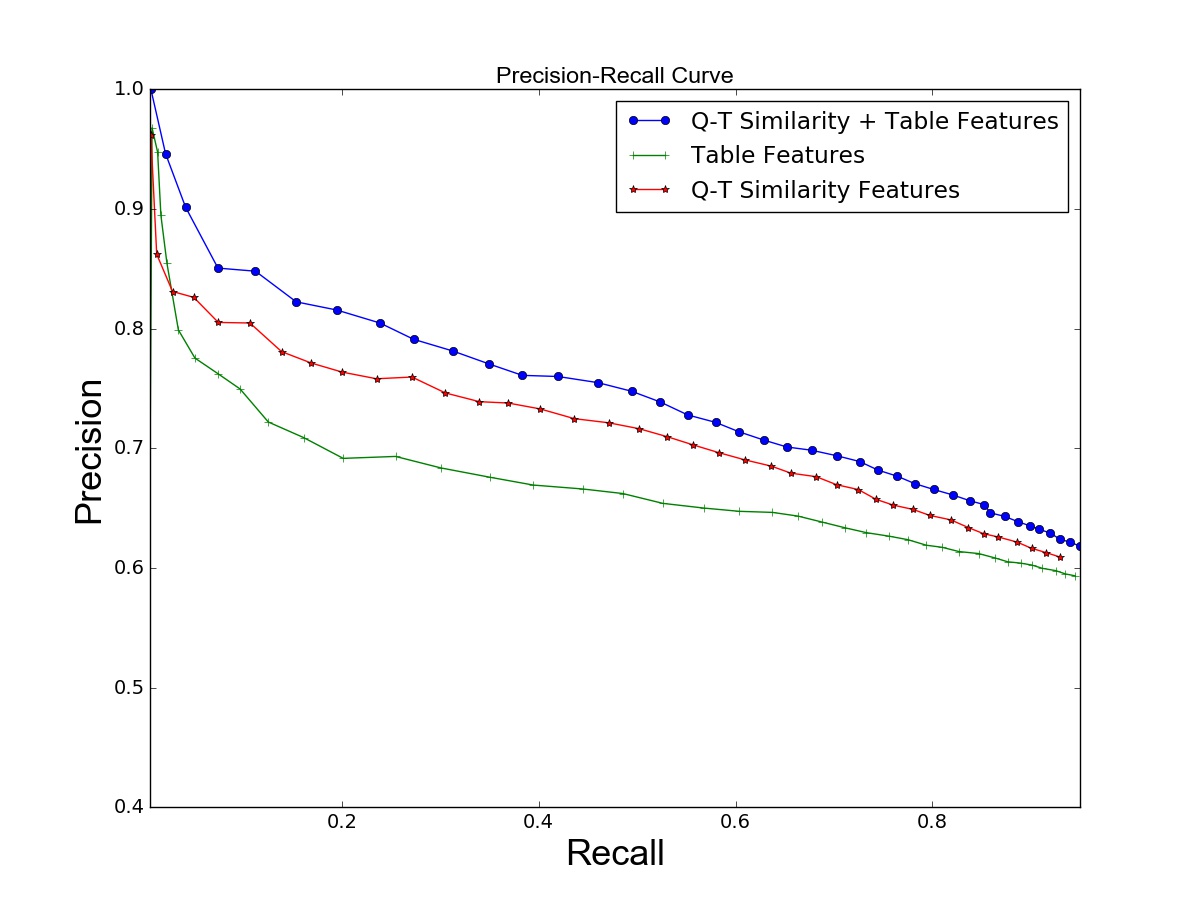}
  \vspace{-0.15in}
 \caption{\small Comparison of different approaches wrt selection performance} \label{fig:CompareApproaches2}
\end{minipage}
\vspace{-0.1in}
\end{figure*}

\begin{figure*}
\vspace{-0.06in}
\begin{minipage}{0.47\linewidth}
\centering
 \includegraphics[width = 3.2in,clip]{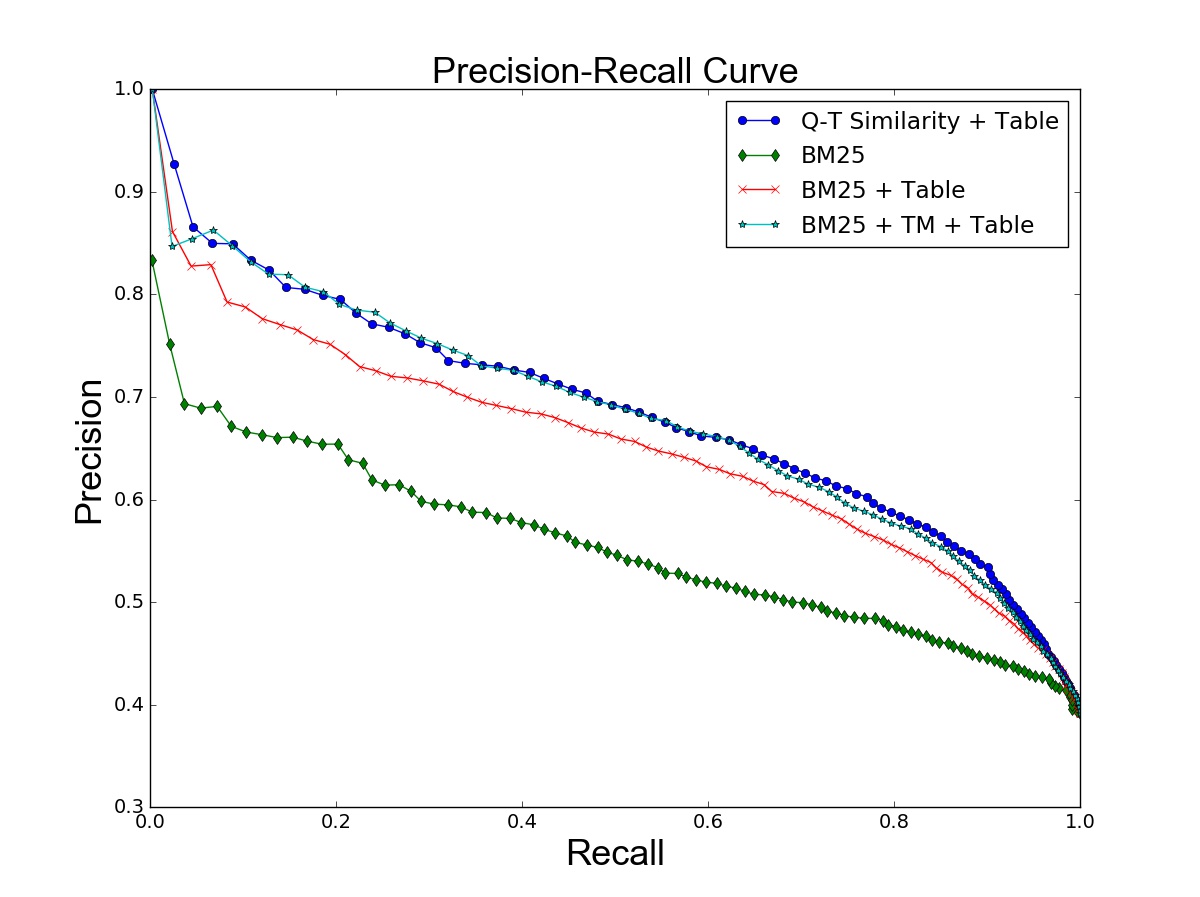}
  \vspace{-0.15in}
 \caption{\small Comparison of query-document similarity measures} \label{fig:SemanticOverBM25}
\end{minipage}
\hspace{0.05\linewidth}
\begin{minipage}{0.47\linewidth}
\centering
 \includegraphics[width = 3.2in,clip]{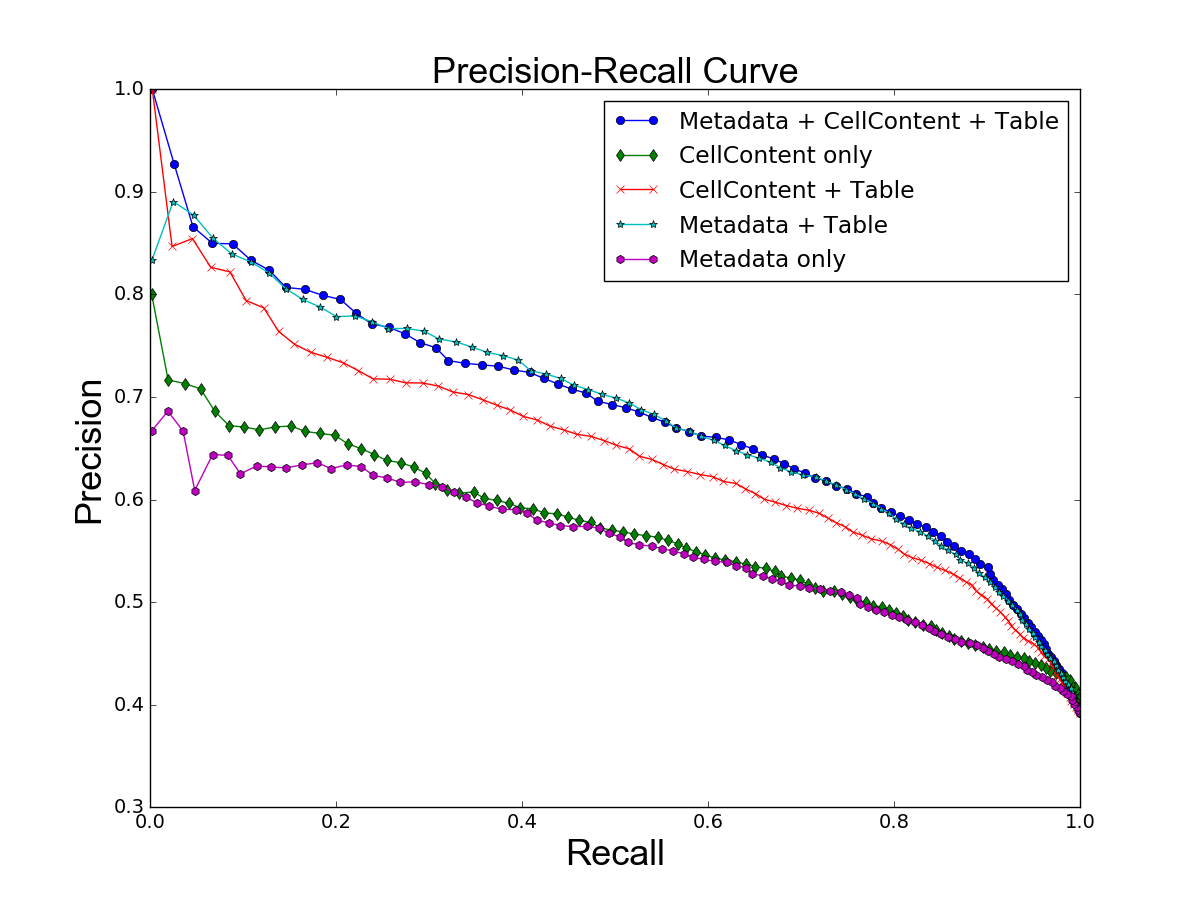}
 \vspace{-0.15in}
 \caption{\small Comparison of content used to construct documents} \label{fig:MetadataAndContent}
\end{minipage}
\vspace{-0.1in}
\end{figure*}

\noindent \textbf{Evaluation Measures:}
We evaluate both the table answer classifier
as well as the table answer selector.
We first consider the table answer classifier.
We compute the precision and recall for a particular threshold $\alpha$ as follows.
For each Q-T pair in the test set, we invoke the classifier, obtain the score $F(Q,T)$,
compare it with $\alpha$ and mark it as one of the following: \\
\noindent $\bullet$ true positive if $F(Q,T) \geq \alpha$ and true label is 1\\
\noindent $\bullet$ false positive if $F(Q,T) \geq \alpha$ and true label is 0 \\
\noindent $\bullet$ false negative if $F(Q,T) < \alpha$ and true label is 1. \\
Precision is $\frac{tp}{tp + fp}$
and recall is $\frac{tp}{tp + fn}$
where $tp$, $fp$ and $fn$ denote the number of true positives,
false positives and false negatives respectively.

We now consider the table answer selector.
We again compute the precision and recall for a particular threshold $\theta$ 
but the computation is slightly different.
For each query, we invoke the table answer selection
algorithm and mark it as one of the following: \\
\noindent $\bullet$ true positive if it returns a result and returned Q-T pair has a true label of 1 \\
\noindent $\bullet$ false positive if it returns a result but returned Q-T pair has a true label of 0 \\
\noindent $\bullet$ false negative if it did not return a result but the query has at least one Q-T pair with true label 1 \\
We compute precision and recall using the same formula as above
but based on this new definition of true positive, false positive and false negative.


\noindent \textbf{Techniques compared:} Since there is no prior work on question answering
using web tables that works for both factoid and non-factoid queries,
there is no prior technique to compare with.
We compare the techniques proposed in
this paper along different dimensions as summarized in Table \ref{tab:variants}.
We compare the techniques along one dimension at a time.
We use the default techniques along all the other dimensions; those are 
are highlighted in bold in Table \ref{tab:variants}.
For example, when we compare the various query-doc similarity measures,
we use the `Q-T Similarity+Table' approach, `Metadata+CellContent' for content, 
`MDoc+CDoc' query-to-doc mapping and `all' table features.

\subsection{Experimental Results}
\noindent \textbf{Comparison of approaches:}
We compare the following 3 approaches wrt to the classifier performance:
the proposed
approach of using Q-T similarity features
in conjunction with table features,
the one using only Q-T similarity features and the one using only table features.
Note that the latter two approaches correspond to the two baselines described in Section \ref{sec:intro}, namely
IR-based and dominating table approaches respectively.
Figure \ref{fig:CompareApproaches1}
shows the precision and recall
of the table answer classifier for 3 approaches at various thresholds. 
The proposed approach significantly outperforms both the baselines for all thresholds.
Since table answers need to be of very high precision, we operate
in the part of the curve where the precision is between $0.8$ and $1$.
For example, at $0.8$ precision, the proposed approach achieves \textbf{$20\%$} recall
while the other two has only \textbf{$4\%$} recall. 
At $0.9$ precision, the former achieves \textbf{$7\%$} recall
while the other two has only \textbf{$2\%$} recall. 
This confirms
that the combined approach safeguards us against
the pitfalls of the two baselines.

We next compare the 3 approaches wrt to the selector performance.
Figure \ref{fig:CompareApproaches2} shows
the precision and recall of the selector for the same 3 approaches
at various thresholds. As with the classifier,
the proposed approach significantly outperforms the two baselines.
For example, at $0.8$ precision, the proposed approach achieves $32\%$ recall
while the Q-T similarity only and table only baselines have only $15\%$
and $5\%$ recall respectively.
At $0.9$ precision, the former achieves $8\%$ recall
while the Q-T similarity only and table only baselines have only 
$1\%$ and $3\%$ recall respectively.

We observe a strong correlation between the performance of the table answer selector
and table answer classifier for all the experiments. This is expected since all the ``real work'' is 
performed by the classifier. Hence, to avoid duplication,
we report only the classifier performance in all subsequent experiments.

\begin{figure*}
\vspace{-0.06in}
\begin{minipage}{0.47\linewidth}
\centering
 \includegraphics[width = 3.2in,clip]{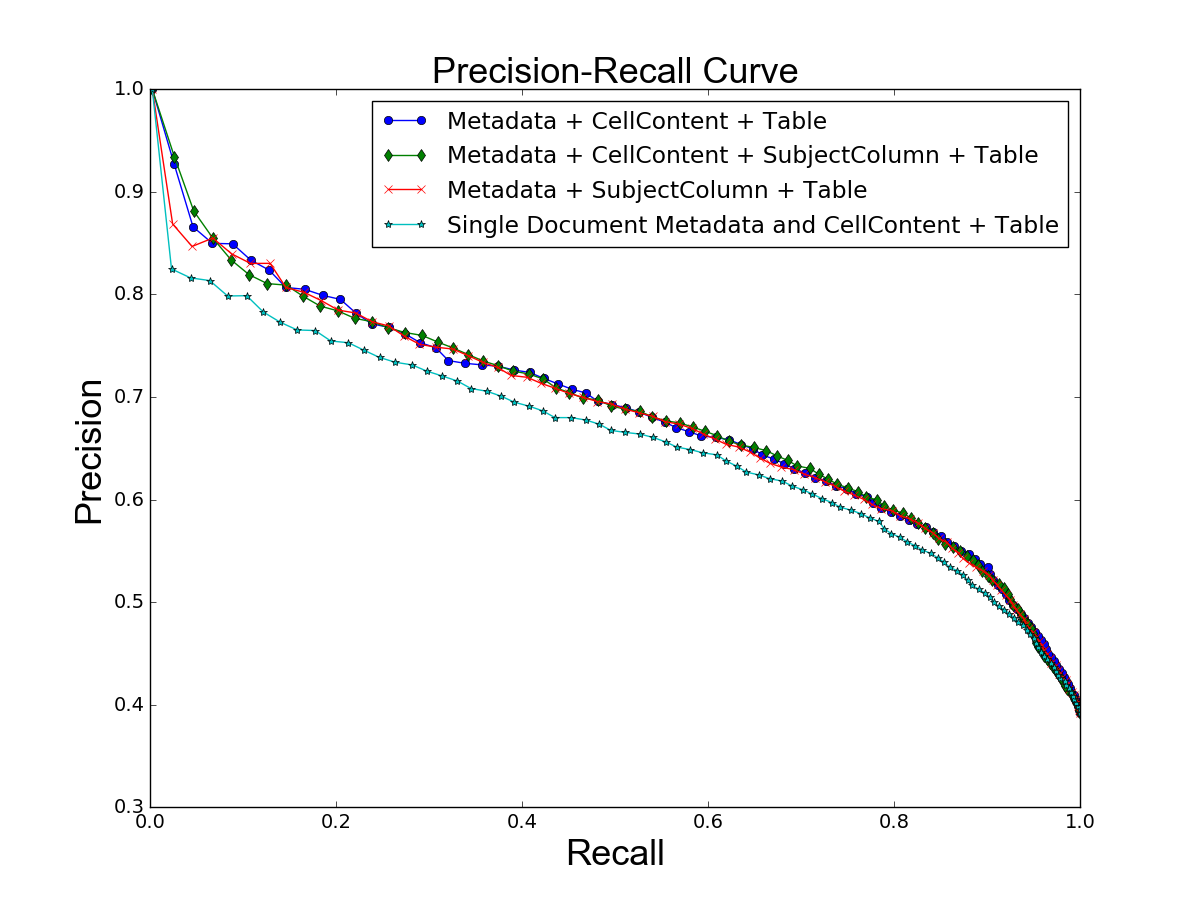}
  \vspace{-0.15in}
 \caption{\small Comparison of table to document mapping techniques} \label{fig:CompareMappingTechniques}
\end{minipage}
\hspace{0.05\linewidth}
\begin{minipage}{0.47\linewidth}
\centering
 \includegraphics[width = 3.2in,clip]{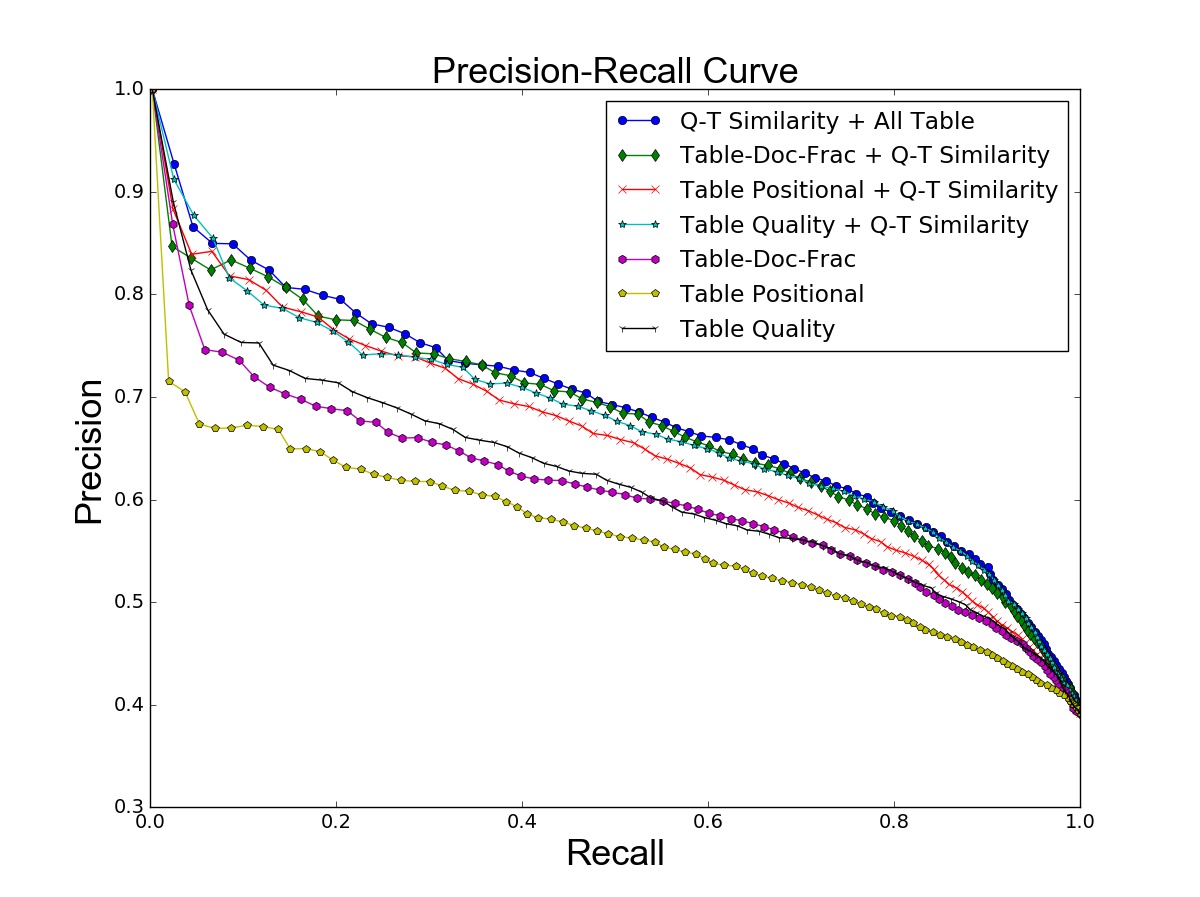}
  \vspace{-0.15in}
 \caption{\small Comparison of groups of table features} \label{fig:CompareTIQGroups}
\end{minipage}
\vspace{-0.1in}
\end{figure*}

\noindent \textbf{Comparison of query-document similarity measures:} 
We compare 3 approaches based on the query-document similarity measured used:
using only BM25, using 
BM25 and one semantic similarity model (translation model)
and  using BM25 and both semantic similarity models (translation model and CDSSM).
As indicated in Table \ref{tab:variants},
the documents are created using the $MDoc+CDoc$ mapping technique.
Figure \ref{fig:SemanticOverBM25}
shows the precision and recall of all the approaches at various thresholds.
BM25 with both semantic similarity models (labeled `All Features')
and BM25 with translation model (labeled as BM25+TM+Table)
significantly outperforms the approach using only BM25 (labeled as BM25+Table).
For example, at $0.8$ precision, the first two
achieves $20\%$ recall
while BM25+Table has only $10\% recall.$
Although `All Features' and BM25+TM+Table appear
to have identical performance, there is a significant difference
for very high precision values. At $0.9$ precision,
former has 10\% recall
while latter has only 5\% recall.
This confirms that the semantic similarities CDSSM and TM indeed bridges
the vocabulary gap between queries and tables
and improves quality.

\noindent \textbf{Comparison of content used to construct documents}
We consider using metadata only, cell content only 
and both metadata and cell content to construct 
documents. The former two approaches construct one document for the table (MDoc and CDoc respectively)
while the last one constructs two documents (MDoc and CDoc).
Figure \ref{fig:MetadataAndContent} compares the above approaches
with and without table features.
Using both metadata and cell content significantly outperforms
the approaches that use metadata only or cell content only.

\noindent \textbf{Comparison of table to document mapping techniques:}
We consider 4 table to document mapping techniques:
single document, MDoc+CDoc, MDoc+SDoc and MDoc+CDoc+SDoc.
Note both metadata and cell content is used in all the approaches.
Table features are also used in all the approaches.
Figure \ref{fig:CompareMappingTechniques} show the precision and recall
curves for the 4 approaches.
MDoc+CDoc and MDoc+CDoc+SDoc 
performs the best. They significantly outperform
the single document and MDoc+SDoc approach.
This confirms that the ability to make the distinction
between matches in metadata and cell content 
improves the quality. We found no significant difference
in the quality of MDoc+CDoc and MDoc+CDoc+SDoc
which indicates creating the subject column document 
does not improve the match between the desired entity type in the query and main entity type in the table.

\noindent \textbf{Comparison of groups of table features}
We evaluate the impact of
the 3 groups of table features:
the ones that compute the fraction of the document occupied by the table,
the position of the table in the page 
and table quality features.
Figure \ref{fig:CompareTIQGroups} compares
the 3 groups with and without Q-T similarity features.
Using all the 3 groups of table features results
in the best performance. 
For example, at precision $0.8$, using all table features (along with Q-T similarity)
achieves $20\%$ recall
while using just the table-doc-frac, table position
and table quality (along with Q-T similarity in all cases) 
has recall $16\%$, $14\%$ and $12\%$ respectively. 

In summary, our experiments demonstrate the superiority of the proposed
approach of combining deep neural network based semantic similarity
with table dominance and quality features over
state-of-the-art baselines.

\section{Related Work}\label{sec:related}

\noindent \textbf{Question answering using web tables}: Our work is most related to previous works on question answering using web tables \cite{sun:www2016,pasupat:acl2015}.
Both works focus on answering factoid questions;
they return the precise cell of a table that answers the question.
However, the settings in the two works are different. In \cite{pasupat:acl2015}, Pasupat and Liang assume that the table containing
the answer to the input question is known beforehand. They develop a semantic parser that
parses the question to a logical form. They also convert the table into a knowledge graph and then
execute the logical form on it to obtain the answer. 
On the other hand, in \cite{sun:www2016}, Sun et. al. do not make that assumption.
Given the question, they find the answer among millions of tables in the corpus.
They construct a unified chain representation of both 
the input question and the table cells and then
find the table cell chain that best matches the question chain.
Our work is more related to \cite{sun:www2016} since
our setting is the same as \cite{sun:www2016}, i.e., the table that answers
the question is not known.
Since our goal is answer both factoid and non-factoid queries,
their techniques cannot be easily applied to our problem.

\noindent \textbf{Question answering using text passages}: 
Researchers have been studying question answering using text passages for several decades
\cite{pascabook:2003,jurafskybook:qachapter,brill:emnlp2002,lin:tois2007}.
They follow the IR-based strategy of matching the question with the candidate 
passages discussed in Section \ref{sec:intro}.
However, as discussed in Section \ref{sec:intro}, this strategy alone is not adequate for returning table answers with high precision. 
This is validated by our experimental results.

\noindent \textbf{Question answering using knowledgebases}:
There is a rich body of work on question answering using knowledge bases \cite{berant:emnlp2013,yih:acl2015,fader:kdd2014,unger:www2012,yahya:emnlp2012}. 
They parse the question into specific forms such as logic forms, graph queries
and SPARQL queries, which can then be executed again the knowledgebase to find the answer.
These parsing techniques cannot be directly applied to solve the table answer selection problem.

\noindent \textbf{Web table search}:
Searching over web tables has been an active area of research over the past decade.
Several search paradigms have been proposed starting with keyword search \cite{cafarella:vldb08,venetis:vldb11},
find related tables \cite{relatedtables_sigmod12}, find tables based on column names \cite{pimplikar:vldb12}
to appending new columns to existing entity lists \cite{infogather_sigmod12,zhang:sigmod13}.
However, these paradigms are different from question answering 
and hence those techniques cannot be directly applied.

\noindent \textbf{Natural language interface to databases}: Finally,
our work is related to works on natural language interfaces to databases \cite{nlidbintro:1995,popescu:iui2003,popescu:coling2004,li:vldb2014}.
They translate natural language questions to SQL queries which
can then be executed against the database. Like works on question answering using knowledgebases,
they focus on semantic parsing and hence cannot be applied to solve the table answer selection problem.

\section{Conclusion}\label{sec:concl} 

In this paper, we developed a query answering approach using web tables
that works for both factoid and non-factoid queries.
Our main insight is to combine deep neural network-based semantic similarity
between the query and the table with features
that quantify the dominance of the table in the document
as well as the quality of the information in the table.
Our experiments demonstrate the superiority of our approach
over state-of-the-art baselines. 

We plan to study question understanding as well as table annotation
approaches to further improve the precision and coverage of our system.
Going beyond relational tables (specifically, attribute value tables \cite{facto:www11})
is also an open challenge.

\bibliographystyle{abbrv}
\bibliography{../BibDatabases/references,../BibDatabases/webtablebib,../BibDatabases/nlidb}

\end{document}